\documentstyle[12pt,proc,epsfig]{article}
\onecolumn
\begin{document}
\onecolumn
\begin{flushright}
TUHEP-TH-00123\\
hep-ph/0201158
\end{flushright}
\begin{center}{\bf PROBING STRONGLY INTERACTING ELECTROWEAK SYMMETRY 
BREAKING MECHANISM AT HIGH ENERGY COLLIDERS}
\end{center}
\baselineskip=16pt
\vspace{0.5cm}
\centerline{\rm YU-PING KUANG}
\baselineskip=13pt
\centerline{\it Department of Physics, Tsinghua University,}
\baselineskip=12pt
\centerline{\it Beijing, 100084, China}
\vspace{0.2cm}
\baselineskip=13pt
\vspace{0.6cm}
\abstract{We sketch some of our recent studies on probing strongly 
interacting electroweak symmetry breaking mechanism at high energy colliders 
such as the CERN LHC and the future $e^+e^-$ linear collider. The study 
includes both model-dependent and model-independent probes.}
\baselineskip=14pt
\section{Introduction}

Despite of the remarkable success that the standard model (SM) has 
passed all the precision tests in the LEP/SLD experiments, the electroweak 
symmetry breaking (EWSB) sector in the SM is not clear yet. So far,
all results of experimental searches for the Higgs boson are negative. The 
recent lower bound for the Higgs boson mass obtained  from the LEP2 
experiments is $m_H>107.7$ GeV \cite{LEP-m_H}. At present, we only know the 
existence of a vacuum expectation value (VEV) $v=246$ GeV which breaks the 
electroweak gauge symmetry, but we do not know if it is just the VEV of the 
elementary SM Higgs boson. We do not even know if there is really a 
Higgs boson below 1 TeV. The unclear EWSB mechanism is a big puzzle in 
particle physics. Since all particle masses come from the VEV $v$, probing the 
EWSB mechanism concerns the understanding of the {\it origin of all particle 
masses}, which is a very radical problem in physics, and is one of the most 
important topics in current high energy physics. Further experimental studies 
on this important problem can be done at future TeV energy colliders
such as the CERN LHC and the future $e^+e^-$ linear collider (LC). 

From the theoretical point of view, there are several unsatisfactory features
in the SM Higgs sector, e.g. there are so many free parameters related
to the Higgs sector, and the Higgs sector suffers from the well-known problems 
of {\it triviality} and {\it unnaturalness} \cite{Chanowitz}. Usually, people 
take the point of view that the present SM theory is only valid up to a 
certain physical energy scale $\Lambda$, and new physics beyond the SM will 
become important above $\Lambda$. Possible new physics are supersymmetry
(SUSY) and dynamical EWSB mechanism concerning new strong interactions, etc. 
So that probing the EWSB mechanism also concerns the discovery of new physics. 

In the SM, the Higgs boson mass $m_H$ is a free parameter related to the
Higgs self-coupling constant $\lambda$ which is not theoretically predictable.
Experimentally, although the Higgs boson is not found, the precision 
electroweak data may give some hint of the Higgs boson mass. 

First of all, it has been pointed out that the best fit of the SM
to the LEP/SLD precision $Z$-pole data requires the Higgs boson mass to be~~
$~m_H=107^{+67}_{-45}$ GeV, and the upper bound of $m_H$ at the 90$\%$ C.L. is 
$m_H<220$ GeV \cite{EL}. This implies that the Higgs boson should be light.

However, we should keep in mind that this analysis is based on the pure SM 
formulae. It has been shown that once new physics effects are considered, the 
bound may change drastically. An analysis in Ref. 4 
shows that if new physics can affect the oblique corrrection parameter $S$,
and if $S$ is taken as an unknown parameter rather than the SM prediction,
the precision data can be well fitted by
$~~S=-0.20^{+0.24}_{-0.17},~~m_H=300^{+690}_{-310}~{\rm GeV},~~
m_t=172.9\pm 4.8~{\rm GeV}$, and $\alpha_s=0.1221\pm 0.0035$.
~We see that the upper bound of $m_H$ in this analysis is close to 1 TeV.

Another interesting analysis was recently given in Ref. 5. Since the Higgs 
boson is not found, the authors consider the possibility that there is no 
undiscovered particles (like the Higgs boson) below $\Lambda\sim $few TeV.
Then, at the LEP energy, the only particles (unphysical) related to the EWSB
mechanism are the would-be Goldstone bosons (GBs). The system of the GBs
with electroweak interactions can be generally described by the 
electroweak chiral Lagrangian (EWCL) \cite{AW} which can be regarded as the 
low energy effective Lagarangian of the fundamental theory of EWSB (see eqs. 
(1)$\--$(3) in Sec. III).
It is shown that the precision $Z$-pole data are only sensitive to the 
$O(p^2)$ terms in the EWCL, which contain two terms related to the oblique 
correction parameters $S$ and $T$. The authors showed that the precision 
$Z$-pole data can be well fitted by the EWCL with the following values
of the parameters: $~~S=-0.13\pm 0.10,~~T=0.13\pm 0.11,
~~\alpha_s(M_Z)=0.119\pm 0.003$. ~This result means that {\it the $Z$-pole 
precision data can be well fitted even without a Higgs boson below the scale 
$\Lambda$}.

Therefore, {\it the precision $Z$-pole data cannot actually tell us whe-\\ther 
there is a light Higgs boson or not}.  

If there is a light Higgs boson, the interactions related to it are weak and 
are perturbatively calculable. The light Higgs will show up as a narrow 
resonance. If the Higgs boson is heavy, or there is no Higgs boson below 1 
TeV, the interactions related to EWSB will be strong and perturbative 
calculation will not work. From the experimental point of view, if the Higgs 
boson is so heavy that its width is comparable to its mass, it will not show 
up as a clear resonance, and the detection is hard. Therefore, in the strongly
interacting EWSB case, both nonperturbative calculation techniques and 
feasible ways of probing the EWSB mechanism should be developed. 

The above analyses show that both the weakly interacting and strongly 
interacting cases should be considered for probing the EWSB mechanism. A 
lot of studies on searching for a light Higgs boson at high energy colliders 
have been made in the literature. Here we briefly present our recent studies 
on probing the strongly interacting EWSB mechanism which contains two kinds of 
studies. Sec. II is on testing specific strongly interacting EWSB models, and 
Sec. III is on a model-independent probe of EWSB mechanism.
\vspace{0.6cm}
\section{Testing Strongly Interacting EWSB Models}

Introducing elementary Higgs field is the simplest but not unique model for 
the EWSB mechanism. The way of completely avoiding {\it triviality} and 
{\it unnaturalness} is to abandon elementary scalar fields and to introduce 
new strong interactions causing certain fermion condensates to break the 
electroweak gauge symmetry. This idea is similar to those in the theory of 
superconductivity and chiral symmetry breaking in QCD. The simplest model 
realizing this idea is the original QCD-like technicolor (TC) model. However, 
such a simple model predicts a too large value of the $S$ parameter and is 
already ruled out by the LEP data. A series of improved models have been 
proposed to overcome the shortcomings of the simplest model, such as the 
Appelquist-Terning one-family model \cite{AT}, topcolor-assisted technicolor 
models \cite{TC2-1}. etc. Topcolor-assisted technicolor theory is an 
attractive improvement, which combines the technicolor and the 
top-condensate ideas. In this theory, the TC dynamics gives rise 
to the masses of the $u,~d,~s,~c,~$ and $b$ quarks and a small portion of the 
top quark mass, while the main part of the top quark mass comes from the 
topcolor dynamics causing the top quark condensate, just as the constituent
quarks acquiring their large dynamical masses from the QCD dynamics causing 
the quark condensates. In this prescription, the TC dynamics does not cause
a large oblique correction parameter $T$ even the mass difference $m_t-m_b$ 
is so large. Improvement of this kind of model is still in progress. 

Most technicolor-type models contain certain pseudo-Goldstone bosons (PGBs)
including {\it technipions} in the techicolor sector and an isospin
triplet {\it top-pions}. Recently, we have shown that the LEP/SLD data of 
$R_b$ constraints the top-pion mass to a few hundred GeV \cite{TC2_Rb-2}. 
These light particles characterizing the phenomenology of the model.
Direct searches for PGBs have been shown to be possible but not easy.

Since the coupling between the top quark and the EWSB sector is strong due to 
the large top quark mass, a feasible way of testing the strongly interacting 
EWSB models is to test the PGB effects in top quark productions at high 
energy colliders. This has been studied by several authors \cite{TLee,EL94}.
We studied these effects in top quark productions at the 
LHC \cite{gg-PGB-tt} and in $\gamma\gamma$ \cite{gaga-PGB-tt} and $e\gamma$ 
\cite{single-top} collisions at the photon collider based on the future LC. We 
took three typical TC models as examples in the study, namely the 
Appelquist-Terning one-family model (model A) \cite{AT} which is not assisted 
by topcolor, the original topcolor-assisted technicolor model (model B) 
\cite{TC2-1}, and the multiscale topcolor-assisted technicolor model (model C) 
\cite{TC2-2}. The studies include calculating the production cross sections 
and $t\bar{t}$ invariant mass distributions which can provide the knowledge of 
whether the PGB effects can be tested and whether the three kinds of models 
can be experimentally distinguished without relying on the details of the 
models.

\vspace{0.6cm}
\begin{figure}[h]
\vspace*{16cm}
\includegraphics{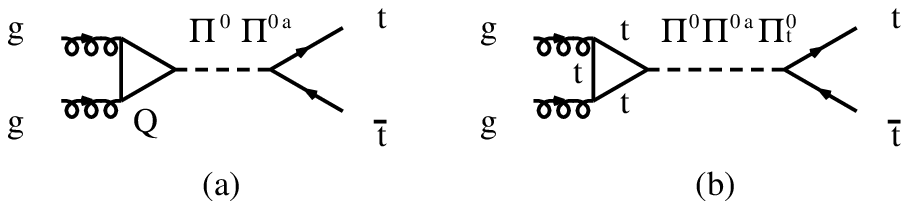}
\vspace{-14.2cm}
\null\noindent{\footnotesize {\bf Fig. 1.}  Feynman diagrams for the 
$s$-channel pseudo-Goldstone boson contributions to the $p(\bar p)\to t\bar 
t$ productions: (a) the techniquark triangle loop contributions, (b) the 
top quark triangle loop contributions. From Ref. 10.}
\end{figure}

At the LHC, there are three kinds of PGBs conributing to the top quark pair
production, namely the neutral color-octet technipion $\Pi^{0a}$ with mass 
around 450 GeV, the neutral color-singlet technipion $\Pi^0$ with mass around
150 GeV, and the color-singlet top-pion with mass in a few hundred GeV region.
Our study in Ref. 10 shows that the $s$-channel PGB contributions
are dominant. 
The PGBs couple to the gluons via the fermion 
triangle loops. For very heavy fermions (techniquarks), the triangle loop was
simply evaluated by the Adler-Bell-Jackiw anomaly. For the top quark triangle 
loop, $m_t$ is not large enough for the validity of simply using the 
Adler-Bell-Jackiw anomaly, and the $m_t$-dependence of this triangle loop was 
taken into account in the calculation. Among the $s$-channel PGB 
contributions, $\Pi^{0a}$ gives the largest resonance contribution to the 
$t\bar{t}$ production cross section. The results in
Ref. 10 show that {\it the PGB effects in the production cross 
sections are large enough to be detected at the LHC}. Furthermore, the
obtained cross sections are quite different for the three typical models. The 
relative differences of the cross sections between model A and model B, model 
A and model C, and model B and model C are $(40\-- 60)\%,~(11\-- 54)\%$, and 
$(11\-- 58)\%$, respectively \cite{gg-PGB-tt}. Considering the statistical 
uncertainty and the expected systematic uncertainty for measuring the 
$t\bar{t}$ cross section at the LHC, {\it the three models can be 
experimentally distinguished at the LHC}. Furthermore, the obtained $t\bar{t}$ 
invriant mass distributions characterized by the resonance peak of $\Pi^{0a}$ 
are quite different in the three kinds of models \cite{gg-PGB-tt}, so
that they are also good observables for distinguishing the three kinds of 
models.

Our results in Ref. 11 for $\gamma\gamma\to t\bar{t}$ and in 
Ref. 12 for $e^+\gamma\to t\bar{b}\bar{\nu}_e$ also lead to the 
same conclusion that the PGB effects in the three kinds of models can be 
detected and they can be distinguished at the LC such as the DESY TESLA.

\vspace{0.6cm}
\section{Model-Independent Probe of EWSB Mechanism}

There have been various kinds of EWSB models proposed, but we do not know 
whether the actual EWSB mechanism in the nature looks like one of them 
or not. Therefore, merely probing the proposed models is not
sufficient, and certain model-independent probe of the EWSB mechanism is 
needed to see what the nature actually looks like. Since the scale of new 
physics is likely to be a few TeV, electroweak physics at energy $E\leq 1$ TeV 
can be effectively described by the EWCL \cite{AW}. The EWCL is a {\it general 
description} (including all kinds of models) which contains certain unknown 
coefficients whose values are determined by the underlying dynamics. Different 
EWSB models give rise to different sets of coefficients. The model-independent 
probe is to investigate through what processes and to what precision we can 
measure these coefficients in the experiments at future high energy colliders. 
From the experimental point of view, the most challenging case of probing
the EWSB mechanism is that there is no light scalar resonance found
below 1 TeV. We took this case as the example in a series of our
investigations \cite{global}\footnote{Effective Lagrangian including a light 
Higgs boson has also been studied in the literature \cite{Buchmuller}.}. In 
the EWCL, the GBs $\pi^a$ with electroweak interactions are 
described in the nonlinear realization $U=e^{i\tau^a\pi^a/f_\pi}$. Up to the 
$p^4$-order, the EWCL reads \cite{AW,global}
\begin{eqnarray}                   
&&{\cal L}_{\rm eff}={\cal L}_{\rm G}+{\cal L}_{\rm S}\,,
\label{CL1}
\end{eqnarray}
where ${\cal L}_{\rm G}$ is the weak gauge boson kinetic energy term, and
\begin{eqnarray}                  
{\cal L}_{\rm S}= {\cal L}^{(2)}+{\cal L}^{(2)\prime}+
             \displaystyle\sum_{n=1}^{14} {\cal L}_n ~~,
\label{CL2}                
\end{eqnarray}
with 
\vspace{0.2cm}
\begin{eqnarray}                  
&&{\cal L}^{(2)}=\frac{f_\pi^2}{4}{\rm Tr}[(D_{\mu}U)^\dagger(D^{\mu}U)],
~~~~~
{\cal L}^{(2)\prime} =\ell_0 (\frac{f_\pi}{\Lambda})^2~\frac{f_\pi^2}{4}
               [ {\rm Tr}({\cal T}{\cal V}_{\mu})]^2 ~~,\nonumber\\
&&{\cal L}_1 = \ell_1 (\frac{f_\pi}{\Lambda})^2~ \frac{gg^\prime}{2}
B_{\mu\nu} {\rm Tr}({\cal T}{\bf W^{\mu\nu}}),\nonumber\\
&&{\cal L}_2 = \ell_2 (\frac{f_\pi}{\Lambda})^2 ~\frac{ig^{\prime}}{2}
B_{\mu\nu} {\rm Tr}({\cal T}[{\cal V}^\mu,{\cal V}^\nu ]) ~~,\nonumber\\
&&{\cal L}_3 = \ell_3 (\frac{f_\pi}{\Lambda})^2 ~ig
{\rm Tr}({\bf W}_{\mu\nu}[{\cal V}^\mu,{\cal V}^{\nu} ]),~~
{\cal L}_4 = \ell_4 (\frac{f_\pi}{\Lambda})^2 
                     [{\rm Tr}({\cal V}_{\mu}{\cal V}_\nu )]^2,\nonumber\\
&&{\cal L}_5 = \ell_5 (\frac{f_\pi}{\Lambda})^2 
                     [{\rm Tr}({\cal V}_{\mu}{\cal V}^\mu )]^2,\nonumber\\
&&{\cal L}_6 = \ell_6 (\frac{f_\pi}{\Lambda})^2 
[{\rm Tr}({\cal V}_{\mu}{\cal V}_\nu )]
{\rm Tr}({\cal T}{\cal V}^\mu){\rm Tr}({\cal T}{\cal V}^\nu),\nonumber\\
&&{\cal L}_7 = \ell_7 (\frac{f_\pi}{\Lambda})^2 
[{\rm Tr}({\cal V}_\mu{\cal V}^\mu )]
{\rm Tr}({\cal T}{\cal V}_\nu){\rm Tr}({\cal T}{\cal V}^\nu),\nonumber\\
&&{\cal L}_8 = \ell_8 (\frac{f_\pi}{\Lambda})^2~\frac{g^2}{4} 
[{\rm Tr}({\cal T}{\bf W}_{\mu\nu} )]^2~,\nonumber\\
&&{\cal L}_9 = \ell_9 (\frac{f_\pi}{\Lambda})^2 ~\frac{ig}{2}
{\rm Tr}({\cal T}{\bf W}_{\mu\nu}){\rm Tr}
        ({\cal T}[{\cal V}^\mu,{\cal V}^\nu ]),\nonumber\\
&&{\cal L}_{10} = \ell_{10} (\frac{f_\pi}{\Lambda})^2\frac{1}{2}
[{\rm Tr}({\cal T}{\cal V}^\mu){\rm Tr}({\cal T}{\cal V}^{\nu})]^2,\nonumber\\
&&{\cal L}_{11} = \ell_{11} (\frac{f_\pi}{\Lambda})^2 
~g\epsilon^{\mu\nu\rho\lambda}
{\rm Tr}({\cal T}{\cal V}_{\mu}){\rm Tr}
({\cal V}_\nu {\bf W}_{\rho\lambda}),\nonumber\\
&&{\cal L}_{12} = \ell_{12}(\frac{f_\pi}{\Lambda})^2 ~2g
                    {\rm Tr}({\cal T}{\cal V}_{\mu}){\rm Tr}
                  ({\cal V}_\nu {\bf W}^{\mu\nu}),\nonumber\\
&&{\cal L}_{13} = \ell_{13}(\frac{f_\pi}{\Lambda})^2~ 
      \frac{gg^\prime}{4}\epsilon^{\mu\nu\rho\lambda}
      B_{\mu\nu} {\rm Tr}({\cal T}{\bf W}_{\rho\lambda}),\nonumber\\
&&{\cal L}_{14} = \ell_{14} (\frac{f_\pi}{\Lambda})^2~\frac{g^2}{8} 
\epsilon^{\mu\nu\rho\lambda}{\rm Tr}({\cal T}{\bf W}_{\mu\nu})
{\rm Tr}({\cal T}{\bf W}_{\rho\lambda}),
\label{CL3}
\end{eqnarray}
\vspace{0.2cm}
in which ${\bf W}_{\mu\nu}$ and $B_{\mu\nu}$ are, respectively, the
field strengths of the gauge fields ${\bf W}_\mu\equiv \tau^aW^a_\mu/2$
and $B_\mu$, $D_{\mu}U =\partial_{\mu}U + ig{\bf W}_{\mu}U 
-ig^{\prime}U{\bf B}_{\mu}~,~~~{\cal V}_{\mu}\equiv (D_{\mu}U)U^\dagger~$, 
~~and $~~{\cal T}\equiv U\tau_3 U^{\dagger}$. 

The coefficients $\ell$s reflect the strengths of the $\pi^a$
interactions, i.e. the EWSB mechanism. $\ell_1,~\ell_0$ and $\ell_8$ are
related to the oblique correction parameters $S,~T$ and $U$, respectively; 
$\ell_2,~\ell_3,~\ell_9$ are related to the triple-gauge-couplings;
${\cal L}_{12},~{\cal L}_{13}$ and ${\cal L}_{14}$ are CP-violating.
The task is to find out experimental processes to measure the  
unknown $\ell$s. 

Note that $\pi^a$ are not physical particles, so that they are not 
experimentally observable. However, due to the Higgs mechanism, the degrees of
freedom of $\pi^a$ are related to the longitudinal components of the weak 
bosons $V^a_L$ ($W^\pm_L,~Z^0_L$) which are experimentally observable. Thus 
$\ell$s can be measured via $V^a_L$-processes. So that we need 
to know the quantitative relation between the $V^a_L$-amplitude (related to 
the experimental data) and the GB-amplitude (reflecting the EWSB mechanism), 
which is the so-called {\it equivalence theorem} (ET). ET has been studied by 
many authors, and the final precise formulation of the ET and its rigorous 
proof are given in our series of papers \cite{et}. The precise form of the
ET is 
\begin{eqnarray}                       
T[V^{a_1}_L,V^{a_2}_L,\cdots]                                   
= C\cdot T[-i\pi^{a_1},\i\pi^{a_2},\cdots]+B ~~,
\label{ET}
\end{eqnarray}
where $~T[V^{a_1}_L,V^{a_2}_L,\cdots]~$ and 
$~T[-i\pi^{a_1},-i\pi^{a_2},\cdots]~$ are, respectively, the $V^a_L$-amplitude 
and the $\pi^a$-amplitude; $C$ is a gauge and renor-\\
malization-scheme dependent 
constant factor, and $B$ is a process-dependent function of the center-of-mass 
energy $E$. By taking a specially convenient renormalization scheme, the 
constant $C$ can be simplified to $C=1$ \cite{et}. In the EWCL theory, the
$B$-term may not be small even when $E\gg M_W$, and {\it is not sensitive to 
the EWSB mechanism}. Therefore the $B$-term serves as an {\it intrinsic 
background} when probing $\pi^a$-amplitude via the $V^a_L$-amplitude in 
(\ref{ET}). Only when $~|B|\ll |C\cdot T[-i\pi^{a_1},-i\pi^{a_2},\cdots]|~$ 
the probe can be {\it sensitive}. 

In our papers \cite{global}, a new {\it power counting rule} 
for semi-quantita-\\tively estimating the amplitudes in the EWCL theory was 
proposed, and with which {\it a systematic analysis on the sensitivities of 
probing the EWSB mechanism via the $V^a_L$ processes at the LHC and the
LC were given}. The results are summarized in the tables 
in Ref. 15. The conclusion is that the coefficients $\ell$s can be 
experimentally determined via various $V^a_L$ processes at various phases of 
the LHC and the LC (including the $e\gamma$ collider) complementarily. Without 
the LC, the LHC itself is not enough for determining all the coefficients. 
Quantitative calculations on the determination of the quartic-$V^a_L$-couplings
$\ell_4$ and $\ell_5$ at the 1.6 TeV LC has been carried out in Ref. 17
which shows that $\ell_4$ and $\ell_5$ can be determined at a higher precision
with polarized electron beams. 
Determination of custodial-symmetry-violating-term coefficients $\ell_6$ and 
$\ell_7$ via the interplay between the $V_LV_L$ fusion and $VVV$ production
has been studied in Ref. 18.

Once the coefficients $\ell_n$s are measured at the LHC and the LC, the
next problem needed to solve is to study what kind of underlying theory will
give rise to this set of coefficients. Only with this theoretical study, the 
systematic probe of the EWSB mechanism can be complete. Such a study is 
difficult due to the nonperturbative nature, and there is no such kind of 
systematic study in the literature.

This kind of study is similar to the problem of deriving the Gasser-Leutwyler
Lagrangian for low lying pseudoscalar mesons (the {\it chiral Lagrangian}) 
\cite{GL} from the fundamental theory of QCD. We can take the QCD case as a 
starting point to do this kind of investigation since the coefficients in the 
Gasser-Leutwyler Lagrangian have already been experimentally measured and
can be compared with the theoretical results. In our recent paper, 
Ref. 20, we developed certain techniques to formally derive the 
Gasser-Leutwyler Lagrangian from the first principles of QCD without taking 
approximations. All the coefficients in the Gasser-Leutwyler Lagrangian 
up to $O(p^4)$ are expressed in terms of certain Green's functions in QCD
\cite{WKWX}, which can be regarded as the {\it QCD definitions of the 
Gasser-Leutwyler coefficients}. The results are \cite{WKWX}:\\

\null\noindent
$O(p^2)$:
\begin{eqnarray}                              
&&F^2_0=\frac{i}{8(N_f^2-1)}\int~d^4x~\bigg[\langle [\bar{\psi}^a(0)\gamma^\mu
\gamma_5\psi^b(0)][\bar{\psi}^b(x)\gamma_\mu\gamma_5\psi^a(x)]\rangle
\nonumber\\
&&\hspace{0.8cm}-\frac{1}{N_f}\langle [\bar{\psi}^a(0)\gamma^\mu\gamma_5
\psi^a(0)]
[\bar{\psi}^b(x)\gamma_\mu\gamma_5\psi^b(x)]\rangle\nonumber\\
&&\hspace{0.8cm}-\langle \bar{\psi}^a(0)\gamma^\mu\gamma_5\psi^b(0)\rangle
\langle\bar{\psi}^b(x)\gamma_\mu\gamma_5\psi^a(x)\rangle\nonumber\\
&&\hspace{0.8cm}+\frac{1}{N_f}\langle\bar{\psi}^a(0)\gamma^\mu\gamma_5
\psi^a(0)\rangle\langle\bar{\psi}^b(x)\gamma_\mu\gamma_5\psi^b(x)\rangle\bigg]
,\nonumber\\
&&F_0^2B_0=-\frac{1}{N_f}\langle\bar{\psi}\psi\rangle.
\label{F_0B_0}
\end{eqnarray}
\null\noindent
$O(p^4)$:
\begin{eqnarray}                          
L_1
&=&\frac{1}{32}{\cal K}_4+\frac{1}{16}{\cal K}_5+\frac{1}{16}{\cal K}_{13}
-\frac{1}{32}{\cal K}_{14}\nonumber,\\
L_2
&=&\frac{1}{16}({\cal K}_4+{\cal K}_6)+\frac{1}{8}{\cal K}_{13}-\frac{1}{16}{\cal K}_{14}
\nonumber,\\
L_3
&=&\frac{1}{16}({\cal K}_3-2{\cal K}_4-6{\cal K}_{13}+3{\cal K}_{14})\nonumber,\\
L_4
&=&\frac{{\cal K}_{12}}{16B_0}\nonumber,~~
L_5
=\frac{{\cal K}_{11}}{16B_0}\nonumber,~~
L_6
=\frac{{\cal K}_8}{16B_0^2}\nonumber,\\
L_7
&=&-\frac{{\cal K}_1}{16N_f}-\frac{{\cal K}_{10}}{16B_0^2}
-\frac{{\cal K}_{15}}{16B_0N_f}\nonumber,\\
L_8
&=&\frac{1}{16}[{\cal K}_1+\frac{1}{B_0^2}{\cal K}_7
-\frac{1}{B_0^2}{\cal K}_9
+\frac{1}{B_0}{\cal K}_{15}]\nonumber,\\
L_9
&=& \frac{1}{8}(4{\cal K}_{13}-{\cal K}_{14})\nonumber,~~
L_{10}
=\frac{1}{2}({\cal K}_2-{\cal K}_{13})\nonumber,\nonumber\\
H_1
&=&-\frac{1}{4}({\cal K}_2+{\cal K}_{13}),\nonumber\\
H_2
&=&\frac{1}{8}[-{\cal K}_1+\frac{1}{B_0^2}{\cal K}_7
+\frac{1}{B_0^2}{\cal K}_9-\frac{1}{B_0}{\cal K}_{15}],\label{p4C}
\end{eqnarray}
where ${\cal K}$s are terms with different Lorentz structures in certain
Greeen's  functions in QCD \cite{WKWX}. Both the anomaly part (from the 
quark functional measure) and the normal part (from the QCD Lagrangian)
give contributions to these coefficients, and the complete coefficients are
given by
\begin{eqnarray}                       
&&F_0^2=(F_0^2)^{(\rm anom)}+(F_0^2)^{(\rm norm)},\nonumber\\
&&F_0^2B_0=(F_0^2B_0)^{(\rm anom)}+(F_0^2B_0)^{(\rm norm)},\nonumber\\
&&L_i=L_i^{(\rm anom)}+L_i^{(\rm norm)},~~~~i=1,\cdots,10,\nonumber\\
&&H_i=H_i^{(\rm anom)}+H_i^{(\rm norm)},~~~~i=1,\cdots,10,
\end{eqnarray}
where the superscripts (anom) and (norm) denote the anomaly part and 
normal part contributions, respectively.
 
In principle, the related QCD Green's functions can be calculated on the 
lattice or in certain approximations. The method in Ref. 20 can be applied 
to the electroweak theory to make the above desired study which is in progress.

\vspace{0.6cm}
\section{Conclusions}

We have briefly sketched the results in several of our recent papers 
concerning the probe of strongly interacting EWSB mechanism at high energy 
colliders such as the CERN LHC and the future $e^+e^-$ linear collider. 
It contains two kinds of studies.
 
In the model-dependent study, we paid special attention to the effects of the
PGBs which characterize the properties of different technicolor models. 
Without relying on the details of the technicolor models, we studied the PGB 
effects in $t\bar{t}$ productions at the LHC, and in $\gamma\gamma$ and 
$e\gamma$ collisions at the $e^+e^-$ linear collider. We showed that not only 
the PGB effects in certain typical improved technicolor models can be detected 
at these high energy colliders, but different typical technicolor models can 
also be experimentally distinguished by measuring the production cross 
sections and the $t\bar{t}$ invariant mass distributions 
\cite{gg-PGB-tt,gaga-PGB-tt,single-top}.
 
In the model-independent study, we have proposed and developed a systematic 
way of measuring the coefficients in the EWCL which relfect the EWSB mechanism,
including the rigorous proof of the equivalence theorem \cite{et}, proposing 
a new power counting rule for estimating the reaction amplitudes in the EWCL 
theory \cite{global}, and a global analysis of the sensitivity of measuring 
the EWCL coefficients at high energy colliders \cite{global}. We showed that 
the EWCL coefficients can be measured at the LHC and the LC, and LHC alone
is not enough for this purpose.
 
We then studied the possiblity of predicting the EWCL coefficients from the 
underlying theory of EWSB mechanism, which may be compared with the measured 
coefficients to get the knowledge of the EWSM mechanism in the nature. This 
kind of study is started from trying to derive the chiral Lagrangian for 
pseudoscalar mesons from QCD, and certain progress has been made in our recent 
paper \cite{WKWX}. The method can be applied to the study of the EWCL, and the 
study is in progress.

\null

\end{document}